\documentclass[a4paper]{article}
\usepackage{epsfig,amsmath,amsfonts,amssymb,setspace,multirow,textcomp}
\usepackage[T1]{fontenc}
\makeatletter
\renewcommand{\@evenfoot}{\hfil \thepage \hfil}
\renewcommand{\@oddfoot}{\hfil \thepage \hfil}
\makeatother

\renewenvironment{thebibliography}[1]{\begin{oldthebibliography}{#1}\setlength{\parskip}{0ex}\setlength{\itemsep}{0ex}}{\end{oldthebibliography}}

\begin{document}
\fontsize{11}{11}\selectfont 
\title{Unidentified Aerial Phenomena. Characterization of Dark UAPs}
\author{Boris Zhilyaev, Vladimir Petukhov, Sergey Pokhvala}
\date{\vspace*{-6ex}}
\maketitle
\begin{center} {\small Main Astronomical Observatory, NAS of Ukraine, Zabalotnoho 27, 03680, Kyiv, Ukraine}\\

{\small bzhi40@gmail.com}

\end{center}

\begin{abstract}
ABSTRACT
We use high-tech observations of Unidentified Aerial Phenomena (UAP) class objects to evaluate their characteristics. We present data in three cases. (1) Multi-side daytime observations of UAPs over Kiev. (2) Night observations of a group of objects in the vicinity of the Moon. (3) UAP observations in the combat zone in Ukraine. Dark UAPs in the visible wavelength range are observed only during the day. At night they can only be seen in the infrared wavelength range. We note large sizes of UAPs, from three to six kilometers.They exhibit large velocities, from 2.5 Mach and much larger. They have low albedo, from three percent and below, that is, they actually exhibit features of a completely black body.


\end{abstract}

\section{Introduction}

The Main Astronomical Observatory of NAS of Ukraine conducts an independent study of unidentified phenomena in the atmosphere. 

Unidentified anomalous, air, and space objects are deeply concealed phenomena. The main feature of the UAP is its high speed.
Ordinary photo and video recordings will not capture the UAP. To detect UAP, we need to fine-tune (tuning) the equipment: shutter speed, frame rate, and dynamic range.

According to our data, there are two types of UAP, which we conventionally call: (1) Cosmics, and (2) Phantoms. We note that Cosmics are luminous objects, brighter than the background of the sky. Phantoms are dark objects, with a contrast, according to our data, to several per cent. Both types of UAPs exhibit high movement speeds. Their detection is a difficult experimental problem. 

The results of previous UAP study are published in \cite{Zhilyaev2022}. Here we present our recent results.

Kyiv astronomers have identified three groups of objects (1) a group of bright spinning objects, (2) a group of bright structured objects and (3) a group of dark flying objects. 
The detection of these objects is an experimental fact.


\section{Analysis of object over Kyiv}

For UAP observations of the objects, we used two meteor stations installed in Kyiv and in the Vinarivka village in the south of the Kyiv region at a base of 120 km. The Vinarivka station has an ASI 294 MC Pro camera and lens with a focal length of 50 mm.  ASI 294 MC Pro camera has a FOV (field of view) of up to 9.7 deg, a pixel size of 34.1 arc second, and a frame rate of up to 120 fps.

The ASI 174 MM camera in Kyiv has an FOV  of 4.08 deg, a pixel size of 24.2 arc second, actual frame rate more of than 200 fps. From simple trigonometry, it is easy to determine that objects at a distance of more than 995 km will fall into the field of view of the cameras at a base of 120 km.

The SharpCap 4.0 program was used for data recording. Observations of objects were carried out in the daytime sky. Frames were recorded in the .ser format with 14 bits. 


Figures 1 and 2 show an image with the object taken synchronously by two cameras in UTC: 17/10/2022, 08:58:08.136 with time precision of one millisecond. 

A parallax of 0.0464 rad (2.66 degrees) gives a distance to the object of 2600 km. For an hight of 26 degrees (Stellarium, UTC: 17/10/2022, 08:58:08) we estimate the altitude of the object at 1130 km.
Vinarivka with 6 shots gives an angular velocity of 1.73 deg/s and a linear velocity of 78 km/s.

An image of the object can be seen in Fig. 3. Fig. 4 shows the color map of the object in RGB rays. The object size can be estimated to be 7 $\pm$ 1 pixels. The PSF (point spread function) of the camera is about 2 x 2 pixels. This gives reason to consider the object as a finite-size object, and estimate the size of the object to be 3 $\pm$ 0.4 km.


\begin{figure}[!h]
\centering
\begin{minipage}[t]{.45\linewidth}
\centering
\epsfig{file = 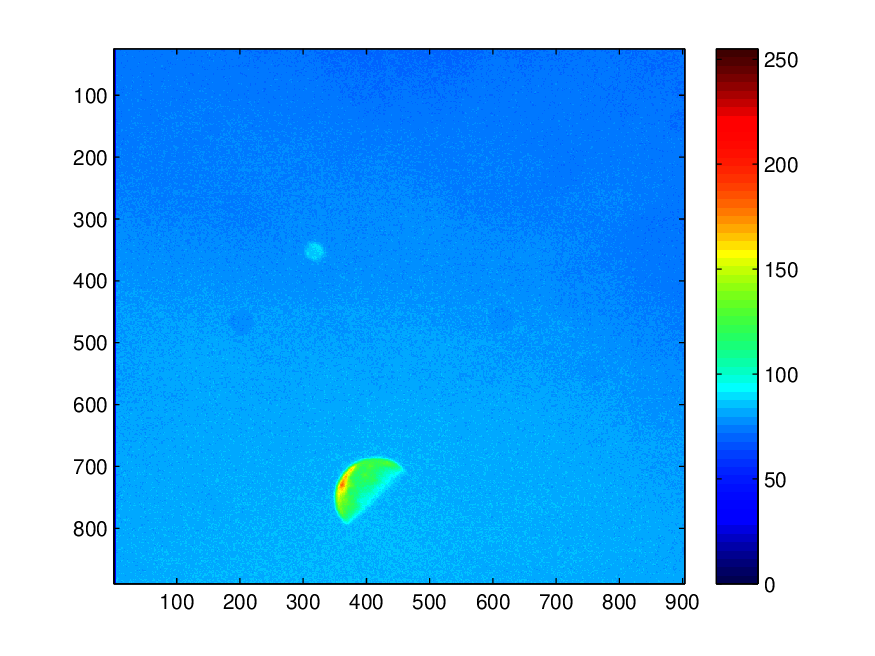,width = 1.05\linewidth} \caption{UAP over Kyiv.}
\end{minipage}
\hfill
\begin{minipage}[t]{.45\linewidth} 
\%\centering
\epsfig{file = 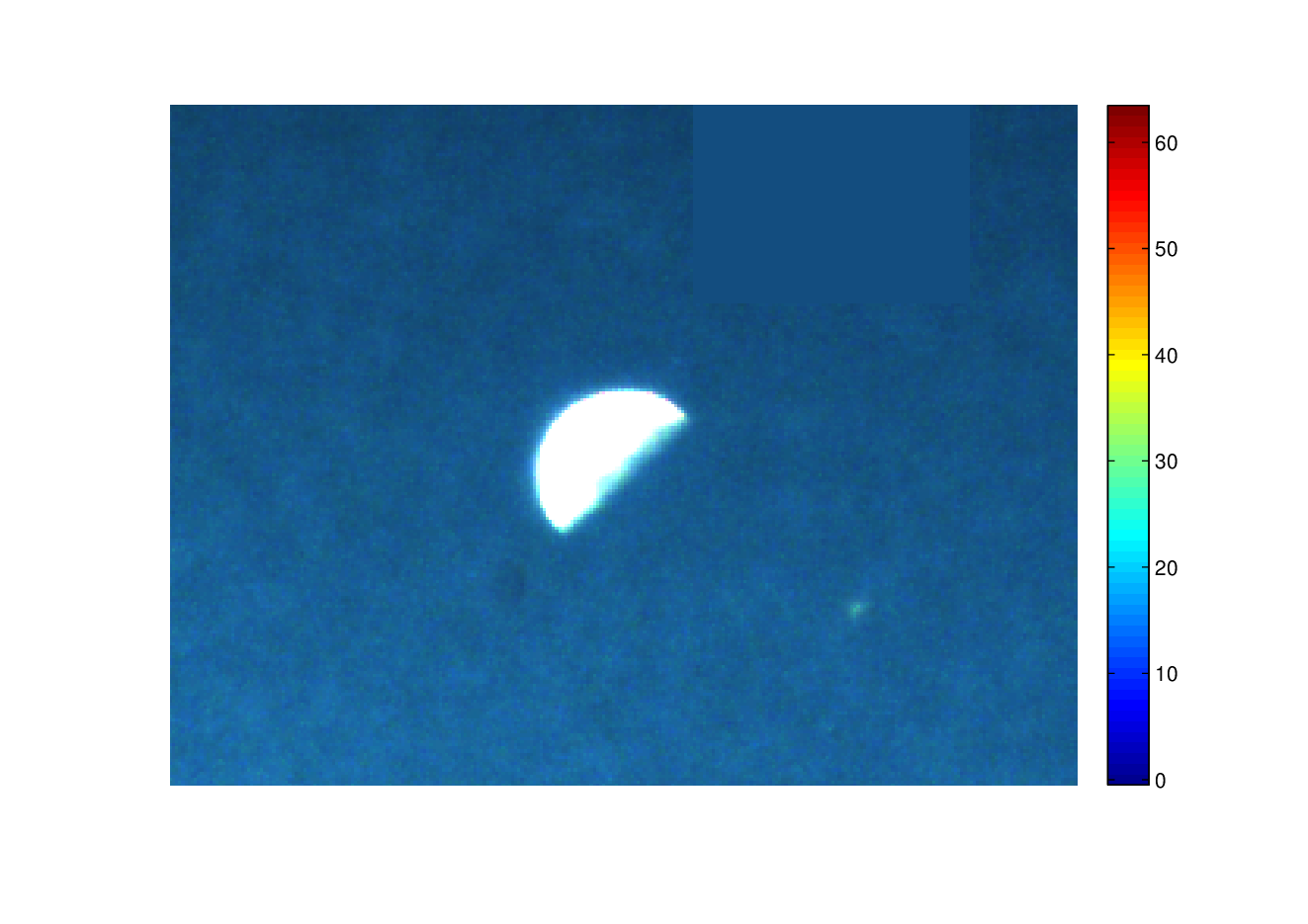,width = 1.05\linewidth} \caption{UAP over Vinarivka.}\label{fig2}
\end{minipage}
\end{figure}

\begin{figure}[!h]
\centering
\begin{minipage}[t]{.45\linewidth}
\centering
\epsfig{file = 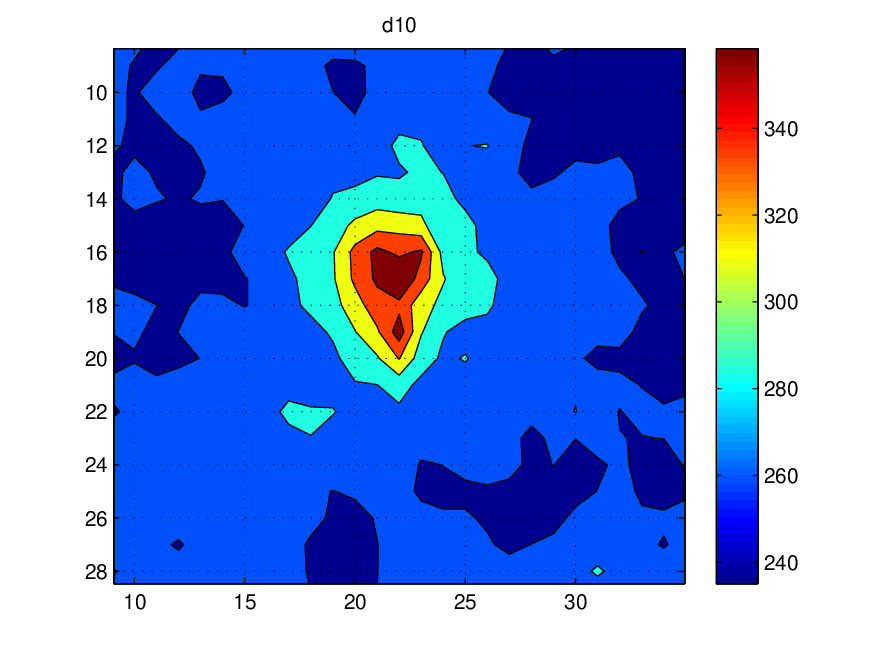,width = 1.05\linewidth} \caption{Image of the object.}
\end{minipage}
\hfill
\begin{minipage}[t]{.45\linewidth} 
\centering
\epsfig{file = 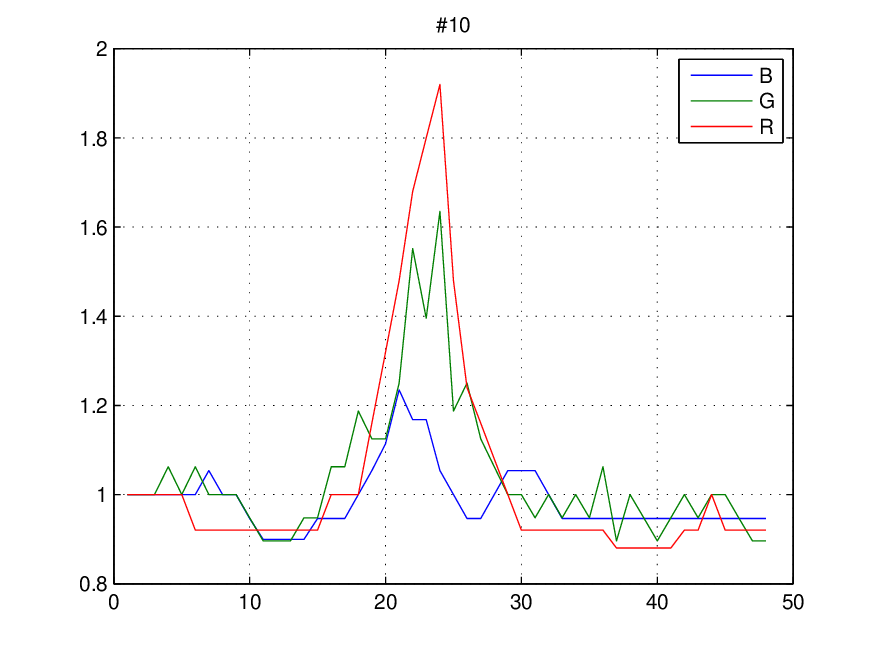,width = 1.05\linewidth} \caption{Color map.}\label{fig4}
\end{minipage}
\end{figure}

\subsection{Evaluation of flying objects' properties}

Observing UAPs, it becomes necessary to evaluate their characteristics. In particular, the sizes of bright objects can be determined if they shine with reflected sunlight. Practice shows that fireflies are visible only in the daytime sky. With the onset of twilight, their brightness decreases, and with the setting of the sun, they disappear from view. If their distance is known from parallax measurements, and if their albedo value is assumed, then one can easily determine the size of the object without angular resolution.

For calculations, we need to know the brightness of the daytime sky near the object and the distribution of energy in the spectrum of the Sun. The generic analytical expression of the spectrum of a clear daylight sky is given in \cite{Zagury}. The brightness of a clear blue sky for a zenith distance of 45 deg depending on the wavelength $ \lambda$ and the distribution of energy in the spectrum of the Sun are given in \cite{Allen}. In particular, a clear daylight sky brightness for $\lambda $ = 0.5 $\mu$m $F_{sky}$ = 4.5 $erg/(cm^{2}\cdot s \,\,  \cdot $\AA$ \cdot sr)$. The solar radiation flux $F_{sun}$ = 193 $erg/(cm^{2}\cdot s\,\, \cdot $\AA$)$.

We represent the light fluxes of the object and the sky background as:
\begin{equation}\label{}
I_{obj}=F_{sun} \cdot r^{2}/R^{2} \cdot \alpha
\end{equation}
\begin{equation}\label{}
I_{sky} = F_{sky}\cdot \Omega
\end{equation}
\begin{equation}\label{}
I_{obj} = \beta \cdot I_{sky}
\end{equation}
Here $r$ is the size of the object, $R$ is the distance to the object, $\alpha $ is the albedo, $\beta $ is the ratio of the brightness of the object to the brightness of the sky background, and $ \Omega$ is the solid angle of the pixel. This implies
\begin{equation}\label{}
r=R\cdot \sqrt{\beta\cdot F_{sky}\cdot \Omega / (F_{sun}\cdot \alpha)}
\end{equation}

If the distance and the size of the object are known from parallax measurements, then one can easily determine the albedo value:
\begin{equation}\label{}
\alpha = \beta\cdot F_{sky}\cdot \Omega / F_{sun}\cdot (R / r)^2
\end{equation}
where $ \Omega$ is the solid angle of the object.

Using the distance and size given above and $\beta$ according to Fig. 4 equal to 2, one can to estimate the albedo of the object to be 0.037.

\section{Analysis of objects near the Moon}
On March 26, 2020, a French astronomer Mark Carlotto used a telescope to capture a video showing the moon at night \cite{Carlotto2}. Dr. M. Carlotto is a specialist in digital video analysis of space objects \cite{Carlotto1}. The video shows three objects rising above the Moon's limb, flying across the lunar surface and disappearing in the Moon's shadow.

What is immediately evident is that the objects in the video are large enough and close enough to the moon to be able to cast noticeable shadows.

The fact that these objects are so clearly visible in the video immediately suggests that they are quite large. Using the large Endymion crater shown in the video as a benchmark, the sizes of the objects were determined.
The size of a single object flying over Endymion is about 5 miles long and about 1 to 3 miles wide. The other two objects appear to be comparable in size.

By measuring the displacement of the object between frames of the video, it appears that the object is traveling at about 31 mps. It is traveling more than 30 times faster than if it were in lunar orbit.

A paper was recently published that attempts to prove that the original video is a fake. We have analyzed the video and made calculations that confirm the originality of the video.

We performed a colorimetric analysis of the surface of the Moon (Endymion crater) and the object in Fig. 5. We represent the light fluxes of the object and the moon background as:
$$I_{obj}=F_{sun} \cdot  \alpha_{obj}$$
and
$$I_{moon}=F_{sun} \cdot  \alpha_{moon}$$

We took RGB intensity estimates of the crater and object points.
Ratio of intensities in white light $\beta$ = $I_{moon}/I_{obj}$ = $\alpha_{moon}/\alpha_{obj} $ = 2.65. The Moon's albedo $\alpha_{moon}$ is 0.067 \cite{Allen}. This gives us an estimate of the object's albedo $\alpha_{obj}$ of 0.025.

The size of Endymion crater in Fig. 5 (bottom panel) is 59 miles. This gives a pixel size estimate of 0.151 miles.
The size of an object flying in Fig. 6 can be estimated to be about 4.3 miles long and about 2.2 miles wide or 6.8 and 3.5 km, respectively.

The size of an object flying in Fig. 5 (bottom panel) is 12 pixels or about 1.8 miles or about 3.0 km. 

The distance of the object flying over the surface of the Moon in Fig. 6 is 36 pixels. The object's height above the Moon's surface is 5.4 miles, or 8.7 kilometers.

The color diagram of the object in RGB rays is shown in Fig. 7.

We can argue that our object estimates and Carlotto's estimates agree within measurement error.


\begin{figure}[!h]
\centering
\begin{minipage}[t]{.45\linewidth}
\centering
\epsfig{file = 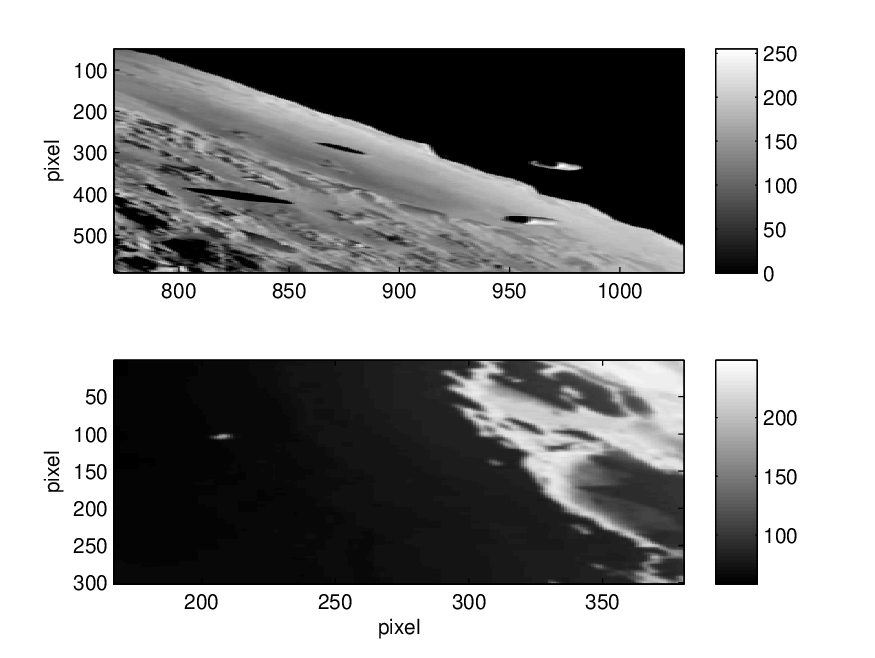,width = 1.05\linewidth} \caption{Frame from original video.}\label{fig1}
\end{minipage}
\hfill
\begin{minipage}[t]{.45\linewidth} 
\centering
\epsfig{file = 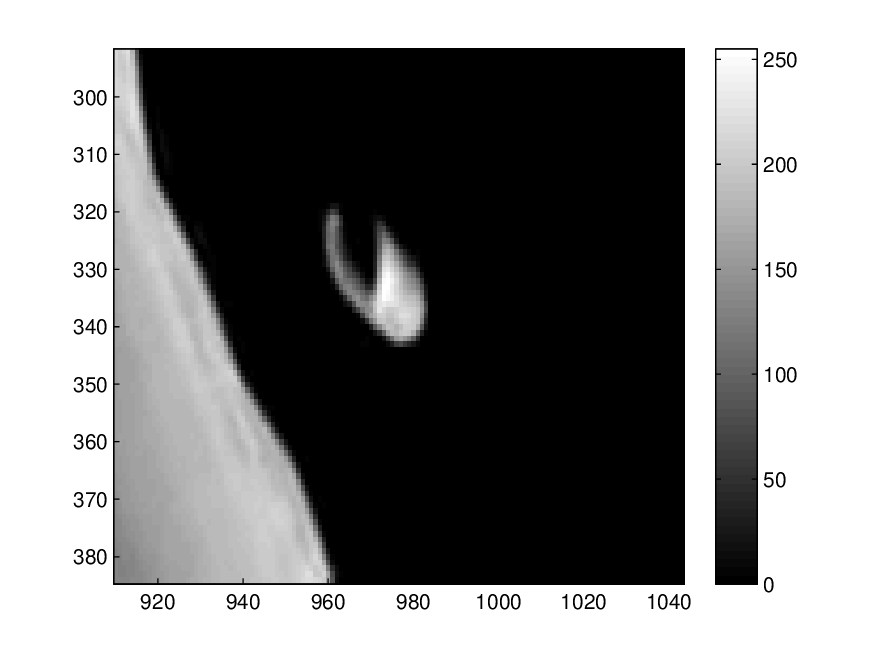,width = 1.0\linewidth} \caption{Frame from original video.}\label{fig2}
\end{minipage}
\end{figure}



\begin{figure}[!h]
\centering
\begin{minipage}[t]{.45\linewidth}
\centering
\epsfig{file = 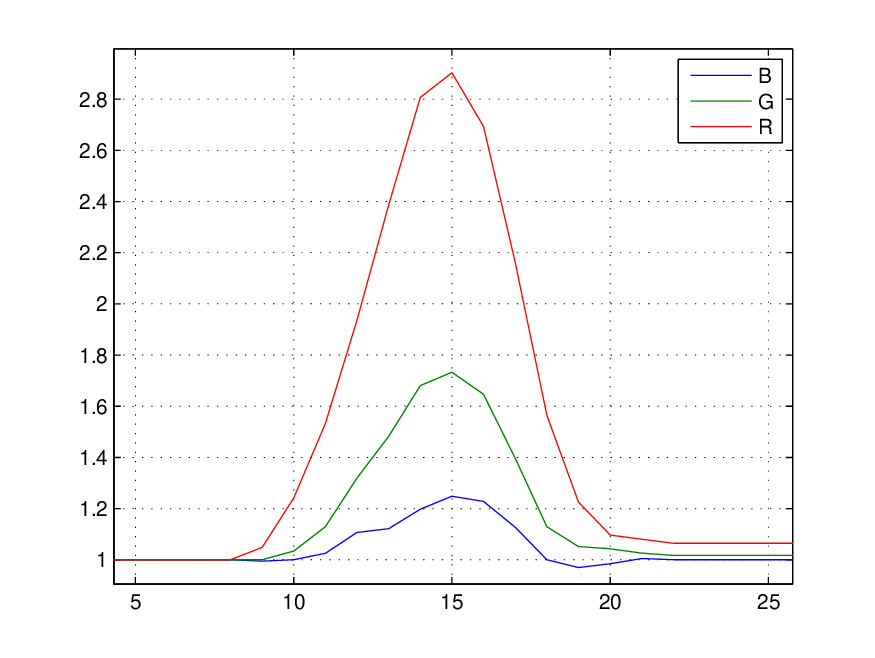,width = 1.05\linewidth} \caption{The color diagram of the object in the RGB Bayer filters.}\label{fig1}
\end{minipage}
\hfill
\begin{minipage}[t]{.45\linewidth} 
\centering
\epsfig{file = 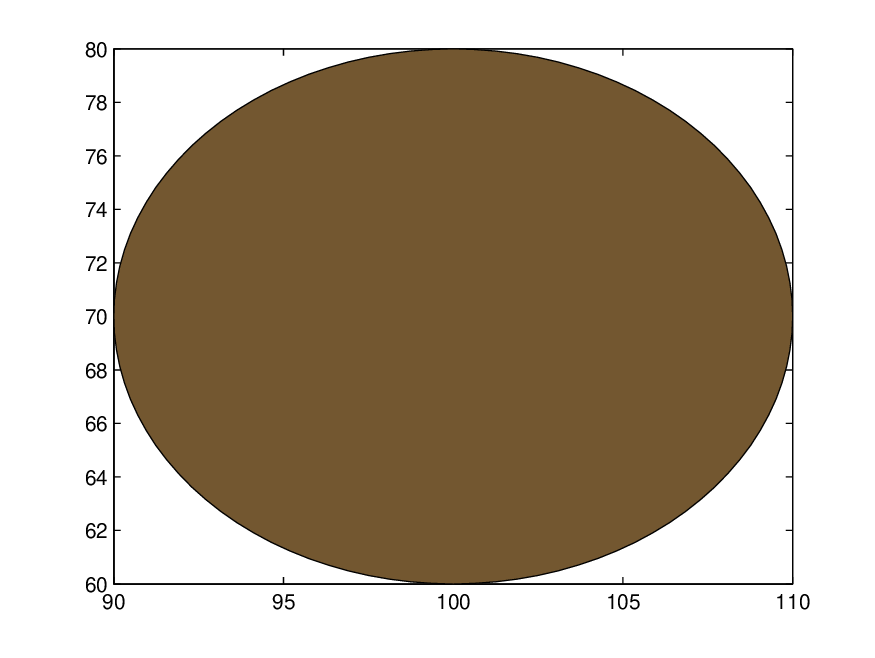,width = 1.0\linewidth} \caption{The restored image of the object.}\label{fig2}
\end{minipage}
\end{figure}


\subsection{Color properties of flying objects}

Fig. 7 shows the color diagram of the object in the RGB Bayer filters. Object colors can be converted to the Johnson BVR astronomical color system using the color corrections published in \cite{Parka}. Semi-empirical relations are as follows:

\begin{equation}\label{}
(B - V )_{J} = 1.47 \cdot (B - G) + 0.12
\end{equation}

\begin{equation}\label{}
(V - R)_{J} = (G - R) + 0.23
\end{equation}

According to Fig. 7 the color index $(B - G)$ is 0.38. Hence $(B - V )_{J}$ is 0.68. Similarly $(V - R)_{J}$ is 0.77. Visually, such an object is perceived as very dark.

For control, we use the formulae (6, 7) to determine the color index $(B - V )_{J}$ in  the Moon. The calculated color index $(B - V )_{J}$ is equal 0.72. According to \cite{Thej}, the measured color index $(B - V )_{J}$  is 0.75 $ \pm$ 0.01.

Using the color chart in Fig. 7, we can restore the color image of the object in the Bayer RGB filters. We use a triple of row vectors $[r g b]$ as indexes to specify the color. Note that the RGB intensity depends on the extinction correction known in \cite{Allen}: [0.91 0.82 0.73]. We have colotmap as
$[rgb]$ = [1/0.91 0.61/0.82 0.43/0.73] $\cdot \,albedo$. The restored image is shown in Fig. 8.


\section{Analysis of the object recorded by the Ukrainian armed forces in the combat zone}

Source of nighttime UAP observations: Instagram account \cite{Inst}. The date of the video upload is 24-2-2024. Video title: "Exclusive footage sent in by a military unit. The 406th battalion of the UAF captured some kind of UAP on its drone while observing the front line".

Later (27-2-2024), the British newspaper Daily Mail wrote about the sighting, also referring to the 406th Battalion video, under the headline "Ukrainian Armed Forces record UFO in a combat zone. EXCLUSIVE: Disk-shaped UFO filmed by Ukrainian military in combat zone."

The video was taken from a DJI Mavic 3T drone with a thermographic camera. The characteristics are given in \cite{Mavik}.

DJI Mavic 3T specifications: (1) sensor - uncooled Vox microbolometer, (2) 12 $\mu$m pixel, (3) 30 Hz frame rate, (4) MP4 video format, (5) infrared wavelength 8 to 14 $\mu$m.

The video is 17 seconds long. A typical frame of the video is shown in Fig. 9.

\begin{figure}[!h]
\centering
\begin{minipage}[t]{.45\linewidth}
\centering
\epsfig{file = 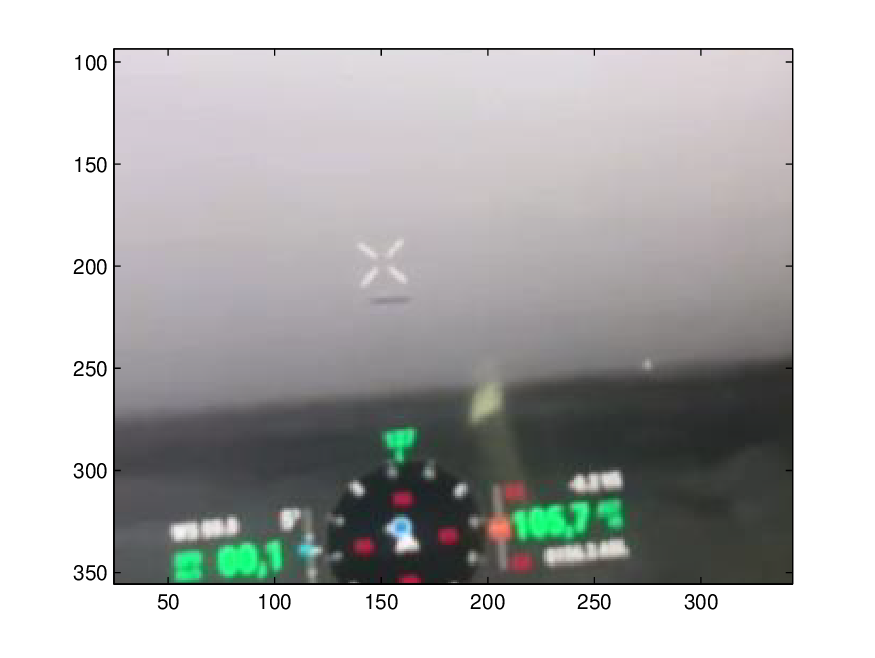,width = 1.05\linewidth} \caption{A typical frame of the video.}\label{fig1}
\end{minipage}
\hfill
\begin{minipage}[t]{.45\linewidth} 
\centering
\epsfig{file = 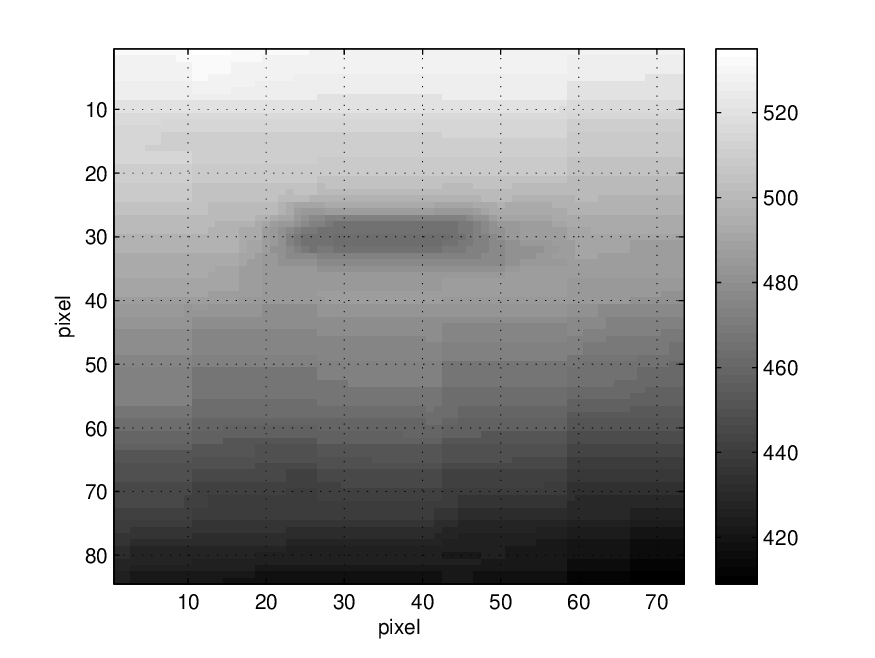,width = 1.0\linewidth} \caption{The object in the video is a dark, elongated body.}\label{fig2}
\end{minipage}
\end{figure}

\subsection{Results}

The object in the video (Figs. 9, 10) is a dark, elongated body, slightly irregular in shape, with a blurred edge.

The Daily Mail article states that "While the size, altitude, and shape of the object remain a mystery, the drone's own altitude indicates that the apparent object could be a large craft over 30 miles away." That is, the expert estimate of the distance is more than 48 km.

Fig. 10 shows a dark object at the end of the observation. It can be stated that the object does not emit and has the characteristics of a completely black body. The low contrast of the image (about 7\% at the beginning of the observation) makes it difficult to detect the object.

The object's brightness and the sky's background allow us to determine the distance by colorimetric methods. A prerequisite is scattering as the main source of atmospheric radiation. The distance can be determined from the residual intensity on the contrast map according to the graph in Fig. 11. 


\begin{figure}[!h]
\centering
\begin{minipage}[t]{.45\linewidth}
\centering
\epsfig{file = 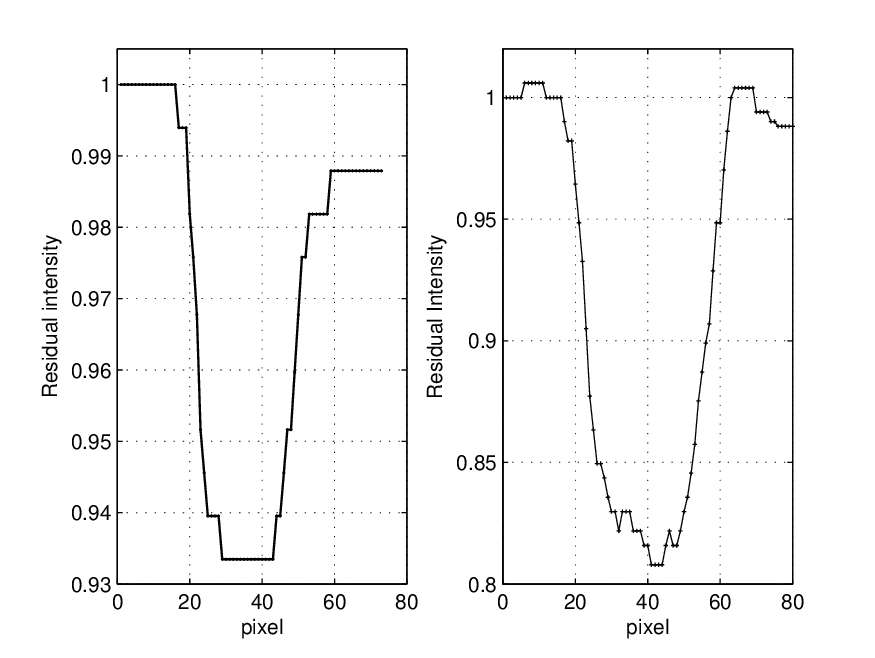,width = 1.05\linewidth} \caption{Contrast map at the beginning of the observation (left panel) and at the end (right panel).}
\end{minipage}
\hfill
\begin{minipage}[t]{.45\linewidth} %
\centering
\epsfig{file = 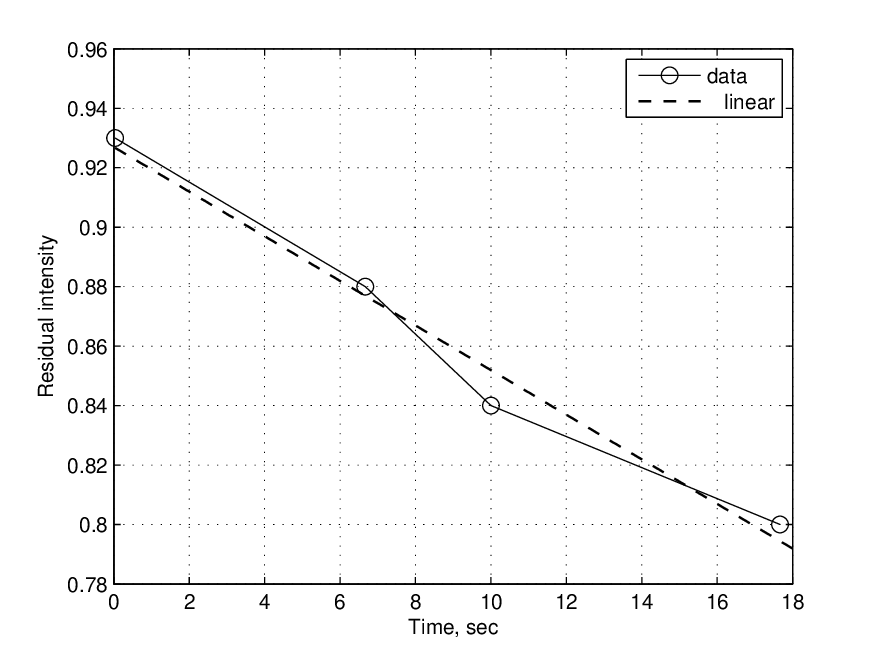,width = 1.05\linewidth} \caption{Linear variations of residual intensity along the trajectory.}\label{fig2}
\end{minipage}
\end{figure}

In the approximation of a homogeneous atmosphere with an altitude of 10 km, the distance $S = 10/sin(h)\cdot r$, where $h$ is the height of the object above the horizon, $r$ is the residual intensity on the contrast map shown in Fig. 11. Using the approximation of a homogeneous atmosphere instead of a real atmosphere with an exponential density distribution gives an error of no more than 6\% \cite{Allen}. 

The height of the object above the horizon at the beginning of the observation is 6.1 degrees. The contrast map of the dark object in the IR wavelength range in Fig. 11 estimates the distance to the object to be 88 km at the beginning and 75 km at the end. The height of the object can be estimated at 8 km.

Fig. 12 shows linear changes of the residual intensity along the trajectory. This indicates that the velocity of the object is constant. Its radial velocity will be about 806 m/s (about 2.5 M).

The angular size of the object can be estimated at 40 pixels and its height at 10 pixels (Figure 10). It can be assumed that the object has a width of about 6 km and a height of about 1.5 km.

\subsection{Conclusion}
The object was discovered by chance during a nighttime drone inspection at 6.1 degrees above the horizon. It was initially located 88 km away, at an altitude of 8 km. During an observation time of 17 seconds, it flew 14 km at a speed of 803 m/s (Mach 2.5). 

The object is Disk-shaped and of tremendous size. It resembles a large floating island in the sky Laputa inhabited by scientists and philosophers in Swift's novel about Gulliver.


\section{Discussion}
The most reliable way to detect UAPs is through two-side observations. For two-side observations, it is necessary to synchronize two cameras with an accuracy of one millisecond. Shots at a rate of at least 50 frames per second in a field of view of 5 degrees at a base of 120 km allowing us to detect objects at a distance above 1000 km.

The object taken synchronously by two cameras with a time precision of one millisecond giving a parallax, distance to the object as well as an angular velocity,   a linear velocity and the altitude of the object. 

We've located the object at altitude of 1130 km. It demonstrates high angular velocities and unprecedentedly high linear speeds of 78 km/s.

Colorimetry showed that the object has an albedo lower than that of the Moon (0.037). Assuming that the object shines with reflected light from the Sun, its size is estimated to be about 3 km. In any case, we can state that the object has a very impressive size.

It is striking that UAPs have been observed in the vicinity of the Moon and in the war zone in Ukraine.
These objects have identical characteristics: about 6 km wide and about 1.5 km high. They fly at an altitude of 8 km with a speed of more than Mach 2.

The objects in the vicinity of the Moon were observed at night in visible light. They were illuminated by the Sun, just like the Moon.

An object in the war zone in Ukraine was observed at night in infrared light at wavelengths of 8 - 14 microns. 

All objects considered in this paper have low albedo, from three percent and below, that is, they actually exhibit features of a completely black body. They have unprecedentedly large sizes, several kilometers, and unprecedentedly high velocities.



\begin{thebibliography}{}

\bibitem{Allen} 
Allen C. W., 1963, Astrophysical Quantities, 2d ed., London, Athlone Press

\bibitem{Carlotto2}
 M.J. Carlotto, https://www.youtube.com/watch?v=ScBx2EwSuDo
 
 \bibitem{Carlotto1}
 M.J. Carlotto, 1995, Journal of Scientijc Exploration, Vol. 9, No. 1 , pp. 45-63  

\bibitem{Inst}
https://www.instagram.com/reel/C3vGfLDr-HB/?igsh=MWZweDV3c3BqczE5Ng

\bibitem{Mavik}
https://alpha-photonics.com/en/produkte/dji-mavic-3-thermal-en/

\bibitem{Parka}
 W. Parka, S. Paka, H. Shimb, et al., 2015, arXiv:1501.04778v3 [astro-ph.IM] 2 Sep 2015

\bibitem{Thej} 
Thejll P., Flynn C., Gleisner H., et al., 2014, A\&A, 563, A38

\bibitem{Zagury}
Zagury F., 2012, The Color of the Sky, Atmospheric and Climate Sciences, 2012, 2, 510-517

\bibitem{Zhilyaev2022}
Zhilyaev B.E., Petukhov V.N., Reshetnyk V.M., 2022, Unidentified aerial phenomena II. Evaluation of UAP properties, DOI:https://doi.org/10.48550/arXiv.2211.17085 



\end{thebibliography}
\end{document}